# Some Measurements of Nullable and Non-Nullable Parameter Declarations in Relation to Software Malleability

William Harrison (Bill.Harrison@cs.tcd.ie), Tim Walsh, Paul Biggar

Department of Computer Science
Trinity College
Dublin 2, Ireland[*]

**Abstract**

The usual advantages put forward for including nullability declarations in the type systems of programming languages are that they improve program reliability or performance. But there is another, entirely different, reason for doing so. In the right context, this information enables the software artifacts we produce, the objects and methods, to exhibit much greater malleability. For declaratively typed languages, we can obtain greater software malleability by extending the model of method call so that assurance of a method's availability can be provided by *any* non-nullable parameter, not simply the target parameter, and by allowing the method's implementation to reside in classes or objects other than the target..

This paper examines the question of whether this hypothetical improvement in software malleability is consistent with existing programming practice by examining the question of the extent to which methods in existing software have multiplicities of non-nullable parameters. The circumstance occurs frequently enough to provide an important reason to introduce declarations of nullability into programming languages.

## 1. Introduction: Malleability, Structural Abstraction, and Recombinance

Malleability is the potential for an artifact to be used in a variety of environmental surroundings, for the artifact to fit in changing circumstances. Malleability is much like reusability, except that instead of characterizing how an artifact can be reused during the development of new artifacts, malleability characterizes how flexibly it can be used without change at deployment-time or when its essential characteristics are needed in finding a service at run-time. Malleability is a key characteristic that will be required for software designed for emerging post-object-oriented architectures and environments [11].

One way for programming languages to increase the malleability of their artifacts is not to declare types. But for the sake of manageability and defect detection and isolation, the creation of large software systems demands specifications at the boundaries between substitutable elements to be developed independently. This is why much large-scale commercial software is developed in declaratively typed languages. For the remainder of this paper, we refer only to declaratively typed languages.

Object-oriented programming languages achieved an improvement in malleability over earlier programming languages through the use of subtype polymorphism. They accrue an even greater improvement when they discourage class types in favor of interface types. But the conventional model of method call restricts malleability by insisting that the client be structured to know in which kind of object a function is implemented, and to direct the call specifically to the target object that implements the method. Greater malleability can be obtained through structural typing [13][8][20] and structural abstraction [14], which permits a client's subtyping expectations to differ from those used in an implementation. Structural abstraction is extended to avoid dependence on the structure of service implementations through the use of a broadcast model of method call [11].

---

[*] This work is supported by a grant from Science Foundation, Ireland

A family of languages, called Continuum [23] languages has been proposed to realize structural abstraction in a broadcast model. These languages are upward-compatible extensions of conventional object-oriented languages; and a Java-compatible Continuum language has been explored. Their call model is based on the principle that from a client's point-of-view, *any* of the non-nullable parameters may provide assurance of an available implementation, not just the single target parameter determined in advance. An implementation of a called method can reside in any type of object known to the dispatcher, even if it is not explicitly mentioned as a parameter of the method. Since method implementations need not reside in the particular objects mentioned, the assurances that it is safe to call them can move from one type to another on a dynamic basis, a phenomenon we have called recombinance [10]. Such assurance can be transferred whenever two variables known to be non-nullable are used simultaneously, such as in a method call. Languages with recombinance can provide even greater software malleability than similar languages without.

Improvements in malleability with the Continuum model result from structural typing, structural abstraction, recombinance and a glossary-based approach to method naming and parameter ordering.. We can illustrate them with Figure 1, using the syntax of Continuum/J, which includes Java's syntax as a subset. The example is drawn from the real-life observation that many mobile phones have an option to show the clock on the display, but that on some phones the command to do so is associated with the clock menu and on others it may be with the display menu. In general software, a similar circumstance commonly arises with bidirectional relationships between objects for which either object may be thought of as the anchor for the relationship. In Figure 1, a system written for home use associates the *display* method with Window objects and the *setColor* method with *Clock* objects. This association is made in both client (A) and service provider (A)(a). In the client it is made by using faces called *Window* and *Clock*, respectively. In the server it is made using classes with the same names. Several points should be noted about the figure:

1. A face is a generalization of an interface, in which the object that is usually implicit as the target parameter is instead listed explicitly. The face itself has no implied target. Alternatively, a programmer can use the conventional "interface" as a syntactic sugar. With this sugar, the compiler simply adds the target parameter under the covers.

2. There is additional syntactic sugar being applied even as shown, however. Continuum semantics treat a variable's declaration as having two independent parts. One part is a ***classifier*** that constrains what kinds of objects the variable refers to. Classifiers are nominal tokens, arranged in a declared subtyping hierarchy with "multiple inheritance". The other part is a ***face*** that prevents the assignment of a non-null reference to the variable unless it has been proven that the methods named in the face have implementations known to the dispatcher. Faces form a structural type system with a natural hierarchy formed by set inclusion. The name spaces for classifiers and faces are independent and each may be specified separately in a variable's declaration; but syntactic sugar allows a single name to be specified and used for both, as it is in this example. The use of structural typing for interface description and separated from the nominal class hierarchy provides the first improvement in malleability claimed above.

3. In the home system's service provider (A)(a), the class declaration for Window defines objects that are to be classified as Window and that have an implementation of the *display* method.

4. The client's *clientSetup* method takes two parameters: one classified as a clock and the other as a window. It also requires support for the methods in the *Clock* and *Window* faces. By default in Continuum/J, both of these parameters are "required" – they may not hold null references. The default for local variables is similar, but the default for fields of objects is "optional" because their values represent the state of the object for which a meaningful value may not exist at a given time.

5. The work system was built using a different structure. In it, both the *display* method is implemented by the *Clock* class. With the conventional formulation of method call and type-safety, the client can not use the work server because its association of method implementations with objects is different from the client's. But in a Continuum interpretation, the client can be serviced by the work server. It does not depend on the association of implementations with classes but only on the structural abstraction that a clock object and a window object are available and that a display method has been proven to be available. In the case of the client server, that is implied by the fact that the *Window* reference is not null. In the case of the work server, it is implied by the fact that the *Clock* reference is not null. So, in both cases it is known to be safe to call. The structurally abstract model for method call allows the client to smoothly make use of the differently structured work environment. This provides the second improvement in malleability claimed above.

The sharp-eyed reader may notice that in Figure 1(A)(b), the fact that the availability of *display* has been proven has been shifted from the server's *Clock* to the client's *Window* when *clientSetup* was called, and worry that this is only safe because *display* itself takes both a window and a clock. Figure 1(B) illustrates the phenomenon of recombinance more crisply. In Figure 1(B) we extended the example of Figure 1(A) with the client's expectation that a *setColor* method is associated with clocks, as is the case in the home server of Figure 1(B)(a). However, in the work server, *setColor* is a static method associated with the Window class. There are a few additional points to be noted:

6. In Continuum's model, there is no real difference between static and object methods; the "static" modifier is simply a syntactic note to inhibit the inclusion of an implicit object parameter in the full signature. Such a method might be implemented using a table that associates clocks with their colors.

7. The structural abstraction shared by the client and servers is a simple addition to that of Figure 1(A) – that a clock object and a window object are available and that the *display* and *setColor* methods have been proven to be available.

8. Even though the work server in Figure 1(B)(b) associates the *setColor* method with *Windows*, a successful call to *clientSetup* requires that both *setColor* and *display* have been proven to be available. Since the references to the window and clock are not null, this is easy to establish from their faces within the work server. But entry to clientSetup recombines the known facts and associates them with different reference parameters. In Continuum, methods in the faces of non-nullable parameters can be recombined to support the faces of any other reference parameter. In logic this recombination is equivalent to the proposition that "anything implies a tautology", and by allowing both static and dynamic re-association of methods with references, it provides the third improvement in malleability claimed above.

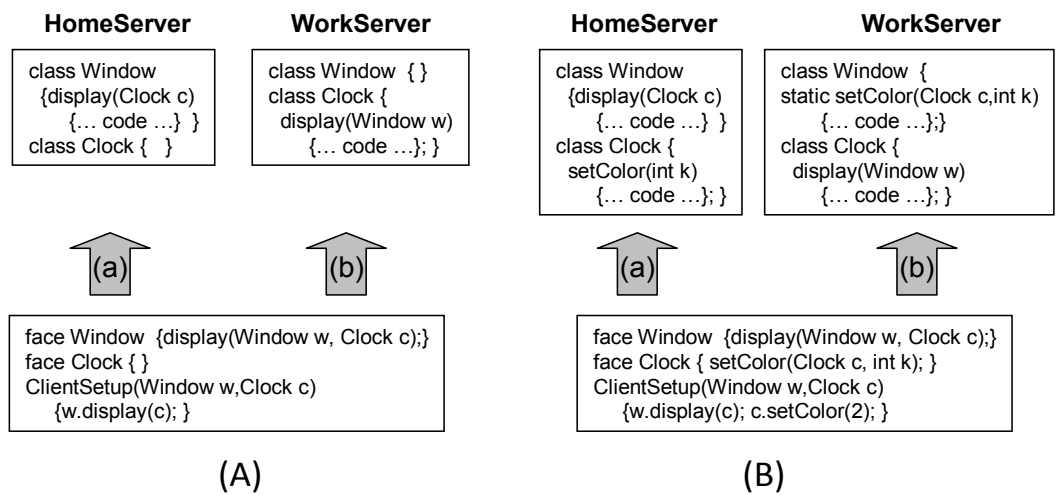

**Figure 1 - Clock client with two servers, structurally abstract and recombinant behavior**

Although the default for reference parameters is that they are required not to be null, providing for the declaration of optional references, preserves the ability to manipulate and use null references. This capability, provided by the common object-oriented languages has been preserved even without the common model's fragility with respect to assignment of methods to a non-null target parameter. We have called the safety model that permits object-oriented languages to treat null references as valid data but to enforce call-safety if they are not null *conditional safety* to contrast it to the unconditional assurances of safety that can be provided with other call models or with the prohibition of null references. The claim that conditional safety is an important characteristic of the method call model is supported by measurements by [2] which indicate that 10% of the parameters to methods in their sample must allow for null values. The dynamic measurements described in Section 3 indicate that the sample used there, 6% of the methods used permit one or more parameter to be null. Exploitation of null values for parameters at this level of utility gives an indication of the importance of dynamic assurance.

## 2. Static Measurements of Potential Occurrence

The clock example described above illustrates the greater malleability to be derived from a recombinant interpretation of method call. But we need to establish that similar situations where recombinance is possible arise in real software. To help evaluate this question we performed a static analysis of some publicly available bodies of software.

We performed a static inter/intra-procedural analysis of the Spec98VM benchmark, containing about 2,000 methods. Each parameter of each method was assigned to one of 3 classifications: 1) possibly required, 2) not locally required, and 3) definitely required. This is done by static analysis of the data flow in the method, starting with the initial assumption that all mentioned variables are not locally required and proceeding iteratively until the information stabilizes. The actual static analysis is performed once to derive a vector function that can be used for the iterations.

Figure 1 illustrates the meanings of the three classifications. The parameter pD is definitely required, because the method unconditionally fails if pD is null. The parameter pP is possibly required because the method might fail if it is null. But if the tested condition implies pI is non-null, it will not fail. And the parameter pN is not locally required because the method will not fail, although if some future execution of another method requires the value to be non-null, the overall program will fail. But if a parameter like pN is passed to a method for which its use is definitely or possibly required, its classification is made more specific. This cross-procedural analysis is performed in the iterative phase.

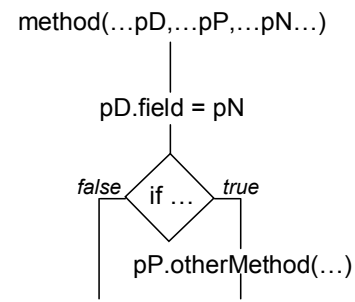

**Figure 1 - Measured Classifications**

The resulting measurements therefore establish a lower-bound on the true proportion of parameters that are actually required in the sample. Table 1 shows, for methods with a given number of reference parameters, how many methods there were in all and what percentage of those are definitely required.

Methods with only one reference parameter other than the target are not susceptible only to what we might call *trivial recombinance*. In our sample, that 56% of all methods in this sample are of this nature and subject only to trivial recombinance. However, although the recombinance is trivial, the capabilities it provides are not. The ability of a programming language to augment the functionality of a class even with respect to such simple methods is highly desirable. Called "obliviousness" [5], this ability is one of the characteristics that defines subject-oriented and aspect-oriented programming approaches to software. The mechanisms for doing so have been called "subjects of an object" [9], "introductions" [21], or "open classes" [19]; and the methods themselves may have static implementations or implementations outside the class entirely. For example, a singleton class "Coloration" can add color attributes, with "get/set" methods to all "Vehicle" class objects by maintaining a separate table on the side, using the Vehicle's reference to look up the colors. Although true recombinance is a multi-parameter property, ability to trivially combine additional methods with those associated with the original objects also improves the overall malleability of software because the client need not be aware of where the implementation resides.

| | | % of methods with "definitely required" parameters | | | | | | |
|---|---|---|---|---|---|---|---|---|
| | | # of reference parameters | | | | | | |
| | | 1 | 2 | 3 | 4 | 5 | 6 | 7 |
| # definitely required parameters | 1 | 100% | 79% | 74% | 73% | 82% | 81% | 100% |
| | 2 | | 21% | 24% | 24% | 16% | 13% | 0% |
| | 3 | | | 2% | 3% | 2% | 0% | 0% |
| | 4 | | | | 0% | 0% | 0% | 0% |
| | 5 | | | | | 0% | 0% | 0% |
| | 6 | | | | | | 6% | 0% |
| | 7 | | | | | | | 0% |
| total # of methods | | 1969 | 920 | 397 | 168 | 44 | 16 | 4 |

**Table 1 - Parameters Definitely Required**

The shaded entries of Table 1 represent parameters of methods that offer opportunities for true recombinance to be effective. They constitute about 23% of all methods with more than one reference parameter. This rises to 25% if we include the possibly required parameters, as shown in Table 2, and, as we have noted, constitutes a lower bound on the actual proportion of cases in which recombinance could be applied to methods with more than one reference parameter. We would consider that this level of potential indicates that opportunities for recombinance arise frequently enough in real software to argue that the malleability of real software can be improved by providing a call model like that illustrated by the clock example, above.

| | | % of methods with "possibly required" parameters | | | | | | |
|---|---|---|---|---|---|---|---|---|
| | | # of reference parameters | | | | | | |
| | | 1 | 2 | 3 | 4 | 5 | 6 | 7 |
| # definitely required parameters | 1 | 100% | 76% | 72% | 72% | 77% | 81% | 100% |
| | 2 | | 24% | 24% | 24% | 20% | 13% | 0% |
| | 3 | | | 4% | 4% | 2% | 0% | 0% |
| | 4 | | | | 1% | 0% | 0% | 0% |
| | 5 | | | | | 0% | 0% | 0% |
| | 6 | | | | | | 6% | 0% |
| | 7 | | | | | | | 0% |
| total # of methods | | 1969 | 920 | 397 | 168 | 44 | 16 | 4 |

**Table 2 - Parameters Possibly Required Locally**

## 3. Dynamic measurements

Although the static inter-procedural analysis techniques were very powerful, the static analysis is conservative. To obtain another estimate of the frequency of use of nullable and non-nullable parameters, we performed a dynamic analysis. For this analysis, we executed the Spec98VM benchmarks 200, 201, 202, 209, 213, 228, and 999, recording for each method call both the (name and) signature of the method and whether or not each of its reference parameters was being passed a null value. There were a total of 260,449,515 method calls reaching 710 different method implementations. We reduced the data, combining the information from all calls of the same method abstraction, no matter which implementation was dispatched, and summarizing, for each parameter, the single fact of whether it was even once passed a null reference value. Abstractly, there were 429 different methods called, with polymorphism reaching the 710 different implementations. In only one case did we note a situation where two implementations of the same abstraction appeared to have different characteristics of null reference permissibility. Examining the Java documentation for this case (ClassDefinition.addDependency(ClassDeclaration c)) provided no clue about the intention of the designers in this regard, and we treated the parameter as nullable. Table 3 summarizes the results:

| never null | number of reference parameters | | | | | | | totals |
|---|---|---|---|---|---|---|---|---|
| | 1 | 2 | 3 | 4 | 5 | 6 | 7 | |
| 1 | 100% | 9% | 0% | 0% | 0% | 0% | 0% | |
| 2 | | 91% | 6% | 0% | 0% | 0% | 0% | |
| 3 | | | 94% | 0% | 10% | 0% | 100% | |
| 4 | | | | 100% | 0% | 25% | 0% | |
| 5 | | | | | 90% | 25% | 0% | |
| 6 | | | | | | 50% | 0% | |
| 7 | | | | | | | 0% | |
| #methods | 126 | 202 | 63 | 24 | 10 | 4 | 1 | 429 |
| #with all required | 126 | 184 | 59 | 24 | 9 | 2 | 0 | 404 |
| %optional | | 9% | 6% | 0% | 10% | 50% | 100% | **5.8%** |

**Table 3 - Proportions of Methods with Various Tolerances of Null Parameters**

We can make two observations from these results. First, by this measure, the potential for recombinance appears much higher than the lower-bound suggested by the earlier static analysis; 92% of all methods with reference parameters in addition to their target offer opportunities for true recombinance. Second, optional parameters are not rare, indicating that the object-oriented feature of conditional safety is exploited in a significant number (6%) of cases, increasing with the number of reference parameters to a method.

Unlike the static analysis described above, the dynamic analysis is optimistic. Perhaps on some other run, a null reference would be passed where none had been in the tests we used. To understand the likelihood of this possibility, we can analyze each of the benchmarks separately. The projects varied in size, and not all of them have methods with more than three reference parameters (including the target). So, for simplicity, we examined the number of methods with two reference parameters in which both of them are never null. From the above Table 3, these comprise 91% of the methods with two reference parameters, but by project, we have:

| | Spec98VM | 999 | 201 | 209 | 200 | 202 | 228 | 213 |
|---|---|---|---|---|---|---|---|---|
| project size | | 1 | 12 | 12 | 15 | 93 | 110 | 167 |
| percentage | | 100% | 60% | 100% | 67% | 98% | 89% | 89% |

**Table 4 - By-project Summary of Methods with Two Reference Parameters, Both Required**

The mean and standard deviation of the percentages is 86%±16%. In any case, the percentages of having several never-null parameters are high enough that even if reduced by one or two standard deviations, they still suggest that the opportunities for recombinance are substantial.

## 4. Related Work

### 4.1. Nullability Declaration

Designers of declaratively-typed programming languages providing reference types must address the question of whether the type-system provides declarative information about whether a null pointer value may be accepted as a variable's value. From this point-of-view, languages can allow it in all cases, prevent it in all cases, or provide a declarative tag to indicate whether or not null values are allowed.

Although the mainstream languages allow nulls to be assigned to any variable, the inclusion of a declarative marker has been advocated and included by designers of functional programming languages like Haskell [12] and ML [16], or object-oriented ones like Nice [1], Theta [15], or Moby [6], and its inclusion in an enhanced version of Eiffel [17]. The reason advocated generally are that declaring nullability raises developers' productivity by improving static detection of defects, and especially by preventing defects which will cause failures far-removed from their point of occurrence. As Fähndrich and Leino point out [4]: "The advantages of adding non-null types to a language like C# or Java include: statically checked interface documentation, statically checked object invariants, more precise error detection, performance optimizations, fewer unexpected null reference exceptions." Failures like null-pointer errors are notoriously expensive to fix because of the lack of direct connection between cause and effect.

To address the need for better software without putting the suppliers of major languages to too much trouble, tools have been developed to allow annotation facilitating the declaration of nullable and non-nullable pointers [3]. Unfortunately, however, extra-lingual mechanisms like annotations often place greater burden on developers attempting to exploit them. As described in [2], it would be preferable, for example, to have variables' declarations default to "non-nullable" rather than nullable, so that a developer only exploits "nullable" when it is a conscious intention. (This would have the minor peripheral advantage of avoiding the use of a "negative" term like "non-null" in favor of one that describes the property as an additional characteristic, like "nullable" or "optional".) More significantly, when the annotations remain outside the language they are not available for deeper semantic changes that they permit, like the structural abstraction and recombinance described in Section 1. In addition, integrating the declarative information in the programming language makes possible a defaulting structure that better suits good practice. For example, fields can default to nullable, permitting complex initialization protocols, while parameters or local variables default to non-nullable to encourage the greater flexibility or clarity it provides.

### 4.2. Motivating Language Features

The motivation for our interest in nullity declaratives is to promote software malleability, in particular the combination of dynamic assurance with structural abstraction or fully symmetric method call. As a general rule, languages, like Cecil, with *generic functions* provide a fully symmetric method call, but do this within a static assurance structure. Languages like C++, C#, Java, MultiJava, etc. provide dynamic assurance, but not fully symmetric call, since clients and developers must agree that implementations are found through the class of the distinguished target parameter. These languages all employ the conventional and multi-method formulations for assurance that provide either structural abstraction or conditional safety, rather than the formulation we described in Section 1, which provides both.

Pedagogical languages like the Dubious [18] family have been used to explore the limits of reconciliation between full symmetry of call and modular checks for avoiding ambiguity. Modular type-checking is especially useful in fully open class-search structures, like Java's *classpath* structure. However, even with the limits suggested for them, these search structures remain open to a variety of module inconsistency problems, often collectively termed *classpath hell* [7]. Other structures, like the OSGi [22] are available, and within such contexts, checking can be limited to within a module, to allow for more structurally abstract software. The "service" structure of Continuum languages [11] is one example of a run-time modular structure larger than that of classes and offering a boundary to limit the costs of checking ambiguity.

### 4.3. Nullability Measurements

There have been several other studies of reference nullity, although to somewhat different ends from ours. While we were interested in counting methods, others have counted individual parameters' statistics which are not readily transformable to method-count statistics. The study by Chalin&James [2] established that in samples they examined, 82% of all references could have been declared non-null and that, for parameters, over 90% could be declared non-null for some applications. Counting individual parameters, our static measurements of parameters indicate that at least 64% of them could be declared non-null, including the target parameter, with the dynamic measurements indicating 97% Chalin&James measurements are higher than the lower-bounds we established, but required hand-insertion of declarations and declarations related to field usage. They are much closer to the dynamic measurements we determined. In a similar vein,

Fähndrich, & Leino examined a C# sample of their own, establishing that 99.5% of the parameters could be declared non-null. We might expect that there are stylistic differences involved in the samples selected.

## 5. Conclusion

We demonstrate the fact that a reformulation of the meaning of assurance provides for greater malleability of software artifacts. This fact provides an important reason to introduce declarations of nullability into programming languages. As observed elsewhere [11], when combined with language permitting parameter reordering, it can be done in a way that compatibly extends conventional object-oriented languages like Java or C++. Measurements indicate that the such declarations in real-world samples are applicable to between 25% and 90% of the methods in a sample examined.